\documentstyle[fullpage,12pt]{article}
\newcommand{\hihi}{\\[-7.5mm]\pagebreak[1]}

\renewcommand{\phi}{\varphi}
\renewcommand{\epsilon}{\varepsilon}

\def\ni{\noindent}

\def\ran{\rangle}

\def\loud#1{\noindent{\bf #1 }}
\def\ee{\vspace{2mm}}
\def\a{\alpha}
\def\b{\beta}
\def\d{\delta}

\def\e{\varepsilon}
\def\th{\theta}
\def\qed{$\Box$}
\def\Rot{\mbox{${\rm R_y}$}}
\def\Ph{\mbox{${\rm R_z}$}}
\def\Scal{\mbox{Ph}}
\def\adj{{\scriptsize \dag}}
\def\adjb{{\dag}}
\def\01{\{0,1\}}

\title{Elementary gates for quantum computation}

\author{Adriano Barenco\\{\protect\small\sl Oxford University}\,%
\thanks{\,Clarendon Laboratory, Oxford OX1 3PU, UK;
a.barenco@mildred.physics.ox.ac.uk.} \and
Charles H. Bennett\\{\protect\small\sl IBM Research}\,%
\thanks{\,Yorktown Heights,
New York, NY 10598, USA; bennetc/divince@watson.ibm.com.} \and
Richard Cleve\\{\protect\small\sl University of Calgary}\,%
\thanks{\,Department of Computer Science,
Calgary, Alberta, Canada T2N 1N4; cleve@cpsc.ucalgary.ca.}
\and
David P. DiVincenzo\\{\protect\small\sl
IBM $\sl Research^\adjb $ }\,%
\and
Norman Margolus\\{\protect\small\sl MIT}\,%
\thanks{\,Laboratory for Computer Science, Cambridge MA
02139 USA; nhm@im.lcs.mit.edu.}
\and
Peter Shor\\{\protect\small\sl AT\&T Bell Labs}\,%
\thanks{\,Murray Hill, NJ 07974 USA; shor@research.att.com.}
\and
Tycho Sleator\\{\protect\small\sl New York Univ.}\,%
\thanks{\,Physics Dept., New York, NY 10003 USA;
tycho@sleator.physics.nyu.edu.}
\and
John Smolin\\{\protect\small\sl UCLA}\,%
\thanks{\,Physics Dept.,
Los Angeles, CA 90024; smolin@vesta.physics.ucla.edu. (and IBM Research.)}
\and
Harald Weinfurter\,%
\\{\protect\small\sl Univ. of Innsbruck}\,%
\thanks{\,Inst. for Exptl. Physics, A-6020 Innsbruck, Austria;
harald.weinfurter@uibk.ac.at.}
}

\date{\small submitted to Physical Review A, March 22, 1995 (AC5710)}

\begin{document}

\maketitle
\thispagestyle{empty}
\setlength{\baselineskip}{0.8cm}

\begin{abstract}
We show that a set of gates that consists of all one-bit quantum gates
(U(2)) and the two-bit exclusive-or gate (that maps
Boolean values $(x,y)$ to $(x,x
\oplus y)$) is universal in the sense that all unitary operations on
arbitrarily many bits $n$ (U($2^n$)) can be expressed as compositions of
these gates.  We investigate the number of the above gates required
to implement other gates, such as generalized Deutsch-Toffoli gates,
that apply a specific U(2) transformation to one input bit if and only
if the logical AND of all remaining input bits is satisfied.  These
gates play a central role in many proposed constructions of quantum
computational networks. We derive
upper and lower bounds on the exact number of elementary gates
required to build up a variety of  two- and three-bit quantum gates,
the asymptotic number required for $n$-bit  Deutsch-Toffoli
gates, and make some observations about the number required for
arbitrary $n$-bit unitary operations.
\end{abstract}

\newpage

\section{Background}

It has recently been recognized, after fifty years of using
the paradigms of classical physics (as embodied in the Turing machine)
to build a theory of computation, that quantum physics provides another
paradigm with clearly different and possibly much more
powerful features than established computational theory.  In quantum
computation, the state of the computer is described by a state vector
$\Psi$,
which is a complex linear superposition of all binary states of the
bits $x_m\in \01$:
$$\Psi(t)=\sum_{x \in \{0,1\}^m}\alpha_x|x_1,\ldots,x_m\ran,\ \ \ \sum_x
|\alpha_x|^2=1.$$
The state's evolution in the course of time $t$
is described by a {\it unitary\/} operator $U$ on this vector space,
i.e., a linear transformation which is bijective and length-preserving.
This unitary evolution
on a normalized state vector is known to be the correct physical description
of an isolated system evolving in time according to the laws of quantum
mechanics\cite{Dira}.

Historically, the idea that the quantum mechanics of isolated systems
should be studied as a new formal system for computation arose from the
recognition twenty years ago that computation could be made reversible
within the paradigm of classical physics.  It is possible to perform any
computation in a way that is reversible both {\it logically\/}---i.e., the
computation is a sequence of bijective transformations---and {\it
thermodynamically\/}---the computation could in principle be performed
by a physical apparatus dissipating arbitrarily little
energy~\cite{Ben1}.
A formalism for constructing reversible Turing machines
and reversible gate arrays (i.e., reversible combinational logic) was
developed.  Fredkin and Toffoli\cite{Fred} showed that there exists a
3-bit ``universal gate" for reversible computation, that is, a gate
which, when applied in succession to different triplets of bits in a
gate array, could be used to simulate any arbitrary reversible
computation. (Two-bit gates like NAND which are universal for ordinary
computation are not reversible.) Toffoli's version\cite{Toff} of the
universal reversible gate will figure prominently in the body of this
paper.

Quantum physics is also reversible, because the reverse-time evolution
specified by the unitary operator $U^{-1}=U^\adj$ always exists; as a
consequence, several workers recognized that reversible computation
could be executed within a quantum-mechanical system. Quantum-mechanical
Turing machines~\cite{Beni,Pere}, gate arrays~\cite{Feyn}, and
cellular automata~\cite{Mar1} have been discussed, and physical
realizations of Toffoli's\cite{Slea,Bar2,Cira} and
Fredkin's\cite{Chua1,Yama,Lloy} universal three-bit gates within various
quantum-mechanical physical systems have been proposed.

While reversible computation is contained within quantum mechanics,
it is a small subset:
the time evolution of a classical reversible computer is
described by unitary operators whose matrix elements are only zero or
one --- arbitrary complex numbers are not allowed.  Unitary time
evolution can of course be {\it simulated} by a classical computer
(e.g., an analog optical computer governed by Maxwell's equations)\cite{Reck},
but the dimension of the unitary operator thus attainable is bounded by
the number of classical degrees of freedom---i.e., roughly proportional
to the size of the apparatus.  By contrast a quantum computer with $m$
physical bits (see definition of the state above) can perform unitary
operations in a space of $2^m$ dimensions, exponentially larger than its
physical size.

Deutsch\cite{Deut} introduced a quantum Turing machine intended to generate
and operate on arbitrary superpositions of states, and proposed that,
aside from simulating the evolution of quantum systems more economically
than known classical methods, it might also be able to solve certain
{\it classical\/} problems---i.e., problems with a classical input and
output---faster than on any classical Turing machine.
In a series of artificial settings, with appropriately chosen oracles,
quantum computers were shown to be qualitatively stronger than classical
ones~\cite{DJ,BB2,simon,BV}, culminating in Shor's~\cite{Shor,RMP}
discovery of quantum polynomial time algorithms for two important
natural problems, viz. factoring and discrete logarithm, for which
no polynomial-time
classical algorithm was known.  The search for other such
problems, and the physical question of the feasibility of building a
quantum computer, are major topics of investigation today\cite{Bras}.

The formalism we use for quantum computation, which we call a quantum
``gate array'' was introduced by Deutsch~\cite{Deu2}, who showed that a
simple generalization of the Toffoli gate   (the three-bit gate
$\wedge_2(R_x)$, in the language introduced later in this paper)
suffices as a universal gate for quantum computing.  The
quantum gate array is the natural quantum generalization of acyclic
combinational logic ``circuits'' studied in conventional computational
complexity theory. It consists of quantum ``gates'', interconnected
without fanout or feedback by quantum ``wires''.  The gates have the
same number of inputs as outputs, and a gate of $n$ inputs carries a
unitary operation of the group U($2^n$), i.e., a generalized rotation in
a Hilbert space of dimension $2^n$.  Each wire represents a quantum bit,
or {\it qubit\/}~\cite{Schu1,JS94}, i.e., a quantum system with a
2-dimensional Hilbert space, capable of existing in a superposition of
Boolean states and of being entangled with the states of other qubits.
Where there is no danger of confusion, we will use the term ``bit''
in either the classical or quantum sense.
Just as classical bit strings can represent the discrete states of
arbitrary finite dimensionality, so a string of $n$ qubits
can be used to represent quantum states in any Hilbert space of
dimensionality up to $2^n$.  The analysis of quantum Turing
machines~\cite{BV} is complicated by the fact that not only the data
but also the control variables, e.g., head position, can exist in a
superposition of classical states.  Fortunately, Yao has
shown~\cite{Yao} that acyclic quantum gate arrays can simulate quantum
Turing machines.
Gate arrays are easier to think about, since the
control variables, i.e., the wiring diagram itself and the number of
steps of computation executed so far, can be thought of as classical,
with only the data in the wires being quantum.

Here we derive a series of results which provide new
tools for the building-up of unitary transformations from simple gates.
We build on other recent results which simplify and extend Deutsch's
original discovery\cite{Deu2}
of a three-bit universal quantum logic gate.  As
a consequence of the greater power of quantum computing as a formal
system, there are many more choices for the universal gate than in
classical reversible computing.  In particular, DiVincenzo\cite{Divi}
showed that
two-bit universal quantum gates are also possible; Barenco\cite{Bar}
extended this
to show than almost any two-bit gate (within a certain restricted class)
is universal, and Lloyd\cite{Llo2}
and Deutsch {\it et al.}\cite{Deu3} have shown that in fact almost
any two-bit or $n$-bit ($n\ge 2$) gate is also universal.
A closely related construction for the Fredkin gate has been
given\cite{Chau}.
In the present paper
we take a somewhat different tack, showing that a non-universal, classical
two-bit gate, in conjunction with quantum one-bit gates, is also universal;
we believe that the present work along with the preceding ones cover the
full range of possible repertoires for quantum gate array construction.

With our universal-gate repertoire, we also exhibit
a number of efficient schemes for building up certain classes of $n$-bit
operations with these gates.
A variety of strategies for constructing gate arrays efficiently will
surely be very important for understanding the full power of quantum
mechanics for computation; construction of such efficient schemes have
already proved very useful for understanding the scaling of Shor's
prime factorization\cite{Copp}.
In the present work
we in part
build upon the strategy introduced by Sleator and Weinfurter\cite{Slea},
who exhibited a scheme
for obtaining the Toffoli gate with a sequence of exactly five two-bit
gates.  We find that their approach can be generalized and extended in
a number of ways to obtain more general efficient gate constructions.
Some of the results presented here have no obvious connection with
previous gate-assembly schemes.

We will not touch at all on the great difficulties attendant on the actual
physical realization of a quantum computer --- the problems of error
correction\cite{Bert2} and quantum coherence\cite{Unr,Chua} are very serious
ones.  We refer the reader to \cite{RL} for a comprehensive discussion
of these difficulties.

\section{Introduction}


We begin by introducing some basic ideas and notation.
For any unitary
$$U = \left(
\begin{array}{cc}
u_{00} &  u_{01} \\
u_{10} &  u_{11}
\end{array}
\right),$$
 and $m \in \{0,1,2,\ldots\}$, define the $(m+1)$-bit
($2^{(m+1)}$-dimensional) operator $\wedge_m(U)$ as
$$\wedge_m(U)(|x_1,\ldots,x_m,y\ran) =
\cases{
u_{y0}|x_1,\ldots,x_m,0\rangle + u_{y1}|x_1,\ldots,x_n,1\ran
& if $\bigwedge_{k=1}^m x_k = 1$ \cr
|x_1,\ldots,x_m,y\rangle
& if $\bigwedge_{k=1}^m x_k = 0$, \cr
}$$
for all $x_1, \ldots, x_m, y \in \01$.
(In more ordinary language, $\bigwedge_{k=1}^m x_k$ denotes the AND of
the boolean variables $\{x_k\}$.)
Note that $\wedge_0(U)$ is equated with $U$.
The $2^{(m+1)} \times 2^{(m+1)}$ matrix corresponding to $\wedge_m(U)$ is
$$\left(
\begin{array}{cccccc}
1 &   &        &   &        &         \\
  & 1 &        &   &        &         \\
  &   & \ddots &   &        &         \\
  &   &        & 1 &        &         \\
  &   &        &   & u_{00} &  u_{01} \\
  &   &        &   & u_{10} &  u_{11}
\end{array}
\right)$$
(where the basis states are lexicographically ordered, i.e., $|000\ran,
|001\ran,\ldots, |111\ran$).

When
$$U = \left(
\begin{array}{cc}
0 & 1 \\
1 & 0
\end{array}
\right),$$
$\wedge_m(U)$ is the so-called Toffoli gate\cite{Toff} with $m+1$ input bits,
which maps $|x_1,\ldots,x_m,y\rangle$ to
$|x_1,\ldots,x_m,\left(\bigwedge_{k=1}^m x_k\right) \oplus y\rangle$.
For a general $U$, $\wedge_m(U)$ can be regarded as a generalization
of the Toffoli gate, which, on input $|x_1,\ldots,x_m,y\rangle$,
applies $U$ to $y$ if and only if $\bigwedge_{k=1}^m x_k = 1$.

As shown by one of us \cite{Deu3,Bar}, ``almost any'' single
$\wedge_1(U)$ gate
is universal in the sense that: by successive application of this gate
to pairs of bits in an $n$-bit network, any unitary transformation
may be approximated with arbitrary accuracy.
(It suffices for $U$ to be specified by Euler angles which are not
a rational multiple of $\pi$.)

We show that in some sense this result can be made even simpler,
in that any unitary transformation in a network can always be constructed
out of only the ``classical'' two-bit gate
$\wedge_1{\scriptsize
\left(
\begin{array}{cc}
0 & 1 \\
1 & 0
\end{array}
\right)
}$
along with a set of one-bit operations (of the form $\wedge_0(U)$).
This is a remarkable result from the perspective of classical reversible
computation because it is well known that the classical analogue of this
assertion---which is that all invertible boolean functions can be
implemented with
$\wedge_1{\scriptsize
\left(
\begin{array}{cc}
0 & 1 \\
1 & 0
\end{array}
\right)
}$
and
$\wedge_0{\scriptsize
\left(
\begin{array}{cc}
0 & 1 \\
1 & 0
\end{array}
\right)
}$
gates\cite{foot1} --- is false.
In fact, it is well known
that only a tiny fraction of Boolean functions
(those which are linear with respect to modulo 2 arithmetic)
can be generated with these gates\cite{Cop2}.

We will also exhibit a number of explicit constructions of $\wedge_m(U)$ using
$\wedge_1(U)$, which can all be made polynomial in $m$.
It is well known\cite{Toff} that
the analogous family of constructions in classical
reversible logic which involve building
$\wedge_m{\scriptsize
\left(
\begin{array}{cc}
0 & 1 \\
1 & 0
\end{array}
\right)
}$ from the three-bit Toffoli gate
$\wedge_2{\scriptsize
\left(
\begin{array}{cc}
0 & 1 \\
1 & 0
\end{array}
\right)
}$, is also polynomial in $m$.  We will exhibit one important difference
between the classical and the quantum constructions, however; Toffoli
showed\cite{Toff} that the classical $\wedge_m$'s could not be built without
the presence of some ``work bits" to store intermediate results of the
calculation.  By contrast, we show that the quantum logic gates can always
be constructed with the use of {\it no} workspace
whatsoever.  Similar computations in the classical setting (that use
very few or no work bits) appeared in the work of Cleve\cite{Clev} and
Ben-Or and Cleve\cite{BenO}.  Still, the presence
of a workspace plays an important role in the quantum gate constructions ---
we find that to implement a family of $\wedge_m$ gates exactly, the time
required for our implementation can be reduced from $\Theta(m^2)$ to
$\Theta(m)$ merely by the introduction of {\it one} bit for workspace.

\section{Notation}

We adopt
a version of Feynman's\cite{Feyn} notation to denote $\wedge_m(U)$ gates and
Toffoli gates in quantum networks as follows.

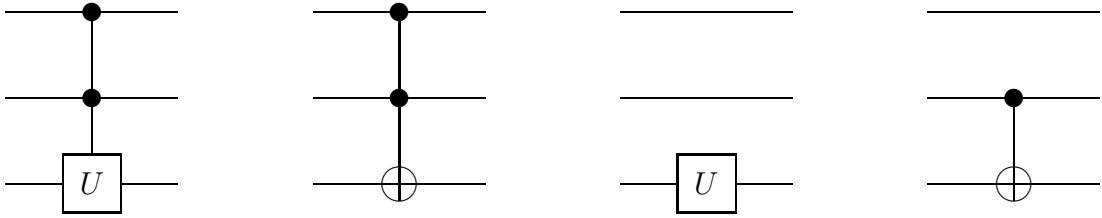
\setlength{\unitlength}{0.030in}

\begin{picture}(30,60)(0,0)

\put(0,15){\line(1,0){10}}
\put(20,15){\line(1,0){10}}
\put(0,30){\line(1,0){30}}
\put(0,45){\line(1,0){30}}

\put(15,30){\circle*{3}}
\put(15,45){\circle*{3}}

\put(15,20){\line(0,1){25}}

\put(10,10){\framebox(10,10){$U$}}

\end{picture}
\begin{picture}(20,60)(0,0)
\end{picture}
\begin{picture}(30,60)(0,0)

\put(0,15){\line(1,0){30}}
\put(0,30){\line(1,0){30}}
\put(0,45){\line(1,0){30}}

\put(15,30){\circle*{3}}
\put(15,45){\circle*{3}}

\put(15,12){\line(0,1){33}}

\put(15,15){\circle{6}}

\end{picture}
\begin{picture}(20,60)(0,0)
\end{picture}
\begin{picture}(30,60)(0,0)

\put(0,15){\line(1,0){10}}
\put(20,15){\line(1,0){10}}
\put(0,30){\line(1,0){30}}
\put(0,45){\line(1,0){30}}

\put(10,10){\framebox(10,10){$U$}}

\end{picture}
\begin{picture}(20,60)(0,0)
\end{picture}
\begin{picture}(30,60)(0,0)

\put(0,15){\line(1,0){30}}
\put(0,30){\line(1,0){30}}
\put(0,45){\line(1,0){30}}

\put(15,30){\circle*{3}}

\put(15,12){\line(0,1){18}}

\put(15,15){\circle{6}}

\end{picture}

\ni In all the gate-array diagrams shown in this paper, time proceeds from
left to right
The first network contains a $\wedge_2(U)$ gate and the second
one contains a 3-bit Toffoli gate\cite{foot2}.
The third and fourth networks contain a $\wedge_0(U)$ and a 2-bit reversible
exclusive-or (simply called XOR henceforth)
gate, respectively.  The XOR gate is introduced as the ``measurement gate"
in \cite{Deu2}, and will play a very prominent role in many of the
constructions we describe below.
Throughout this paper, when we refer to a {\it basic} operation, we
mean either a $\wedge_0(U)$ gate or this 2-bit XOR gate.

In all the gate-array diagrams shown in this paper,
we use the usual convention that time advances from
left to right, so that the left-most gate operates first, etc.

\section{Matrix Properties}

\loud{Lemma 4.1:}{\sl
Every unitary $2 \times 2$ matrix can be expressed as
$$
\left(
\begin{array}{ll}
e^{i \d} & 0 \\
0        & e^{i \d}
\end{array}
\right) \cdot
\left(
\begin{array}{ll}
e^{i \a/2} & 0 \\
0        & e^{-i \a/2}
\end{array}
\right) \cdot
\left(
\begin{array}{rr}
\cos \th/2 & \sin \th/2 \\
-\sin \th/2 & \cos \th/2
\end{array}
\right) \cdot
\left(
\begin{array}{ll}
e^{i \b/2} & 0 \\
0        & e^{-i \b/2}
\end{array}
\right),
$$
where $\d$, $\a$, $\th$, and $\b$ are real-valued.
Moreover, any special unitary $2 \times 2$ matrix
(i.e., with unity determinant) can be expressed as
$$\left(
\begin{array}{ll}
e^{i \a/2} & 0 \\
0        & e^{-i \a/2}
\end{array}
\right) \cdot
\left(
\begin{array}{rr}
\cos \th/2 & \sin \th/2 \\
-\sin \th/2 & \cos \th/2
\end{array}
\right) \cdot
\left(
\begin{array}{ll}
e^{i \b/2} & 0 \\
0        & e^{-i \b/2}
\end{array}
\right).
$$}\ee

\loud{Proof:}
Since a matrix is unitary if and only if its row vectors and column vectors
are orthonormal, every $2 \times 2$ unitary matrix is of the form
$$\left(
\begin{array}{rr}
e^{i (\d + \a/2 + \b/2)} \cos \th/2 & e^{i (\d + \a/2 - \b/2)} \sin \th/2 \\
- e^{i (\d - \a/2 + \b/2)} \sin \th/2 & e^{i (\d - \a/2 - \b/2)} \cos \th/2
\end{array}
\right),$$
where $\d$, $\a$, $\th$, and $\b$ are real-valued.
The first factorization above now follows immediately.
In the case of special unitary matrices, the determinant of the first
matrix must be 1, which implies $e^{i \d} = \pm 1$, so the first matrix
in the product can be absorbed into the second one.\qed\ee

\loud{Definition:}
In view of the above lemma, we define the following.
\begin{itemize}
\item
$\displaystyle{
\Rot(\th) =
\left(
\begin{array}{rr}
\cos \th/2 & \sin \th/2 \\
-\sin \th/2 & \cos \th/2
\end{array}
\right)}$ (a rotation by $\th$ around $\hat y$\cite{Expl}).
\item
$\displaystyle{
\Ph(\a) =
\left(
\begin{array}{ll}
e^{i \a/2} & 0 \\
0        & e^{-i \a/2}
\end{array}
\right)}$ (a rotation by $\alpha$ around $\hat z$).
\item
$\displaystyle{
\Scal(\d) =
\left(
\begin{array}{ll}
e^{i \d} & 0 \\
0        & e^{i \d}
\end{array}
\right)}$ (a phase-shift with respect to $\d$).
\item
$\displaystyle{
\sigma_x =
\left(
\begin{array}{ll}
0 & 1 \\
1 & 0
\end{array}
\right)}$ (a ``negation'', or Pauli matrix).
\item
$\displaystyle{I =
\left(
\begin{array}{ll}
1 & 0 \\
0 & 1
\end{array}
\right)}$ (the identity matrix).
\end{itemize}
\ee

\loud{Lemma 4.2:}{\sl
The following properties hold:}
\begin{enumerate}
\item
$\Rot(\th_1) \cdot \Rot(\th_2) = \Rot(\th_1 + \th_2)$
\item
$\Ph(\a_1) \cdot \Ph(\a_2) = \Ph(\a_1 + \a_2)$
\item
$\Scal(\d_1) \cdot \Scal(\d_2) = \Scal(\d_1 + \d_2)$
\item
$\sigma_x \cdot \sigma_x = I$
\item
$\sigma_x \cdot \Rot(\th) \cdot \sigma_x = \Rot(-\th)$
\item
$\sigma_x \cdot \Ph(\a) \cdot \sigma_x = \Ph(-\a)$
\end{enumerate}
\ee

\loud{Lemma 4.3:}{\sl For any special unitary matrix $W$ ($W\in SU(2)$),
there exist
matrices $A$, $B$, and $C$ $\in SU(2)$ such that
$A \cdot B \cdot C = I$ and $A \cdot \sigma_x \cdot B \cdot \sigma_x \cdot C
= W$.}\ee

\loud{Proof:}
By Lemma 4.1, there exist $\a$, $\th$, and $\b$ such that
$W = \Ph(\a) \cdot \Rot(\th) \cdot \Ph(\b)$.
Set $A = \Ph(\a) \cdot \Rot({\th \over 2})$,
$B = \Rot(-{\th \over 2}) \cdot \Ph(-{\a+\b \over 2})$, and
$C = \Ph({\b - \a \over 2})$.
Then
\begin{eqnarray*}
A \cdot B \cdot C
& = &
\Ph(\a) \cdot \Rot(\textstyle{{\th \over 2}}) \cdot
\Rot(\textstyle{-{\th \over 2}}) \cdot \Ph(\textstyle{-{\a + \b \over 2}})
\cdot
\Ph(\textstyle{{\b - \a \over 2}}) \\
& = &
\Ph(\a) \cdot \Ph(-\a) \\
& = & I,
\end{eqnarray*}
and
\begin{eqnarray*}
A \cdot \sigma_x \cdot B \cdot \sigma_x \cdot C
& = &
\Ph(\a) \cdot \Rot(\textstyle{{\th \over 2}})
\cdot \sigma_x \cdot
\Rot(\textstyle{-{\th \over 2}}) \cdot \Ph(\textstyle{-{\a + \b \over 2}})
\cdot \sigma_x \cdot
\Ph(\textstyle{{\b - \a \over 2}}) \\
& = &
\Ph(\a) \cdot \Rot(\textstyle{{\th \over 2}})
\cdot \sigma_x \cdot
\Rot(\textstyle{-{\th \over 2}})
\cdot \sigma_x \cdot \sigma_x \cdot
\Ph(\textstyle{-{\a + \b \over 2}})
\cdot \sigma_x \cdot
\Ph(\textstyle{{\b - \a \over 2}}) \\
& = &
\Ph(\a) \cdot \Rot(\textstyle{{\th \over 2}})
\cdot
\Rot(\textstyle{{\th \over 2}})
\cdot
\Ph(\textstyle{{\a + \b \over 2}})
\cdot
\Ph(\textstyle{{\b - \a \over 2}}) \\
& = &
\Ph(\a) \cdot \Rot(\th) \cdot \Ph(\b) \\
& = & W.
\end{eqnarray*}
\qed\ee

\section{Two-Bit Networks}

\subsection{Simulation of General $\wedge_1(U)$ Gates}

\loud{Lemma 5.1:}{\sl
For a unitary $2 \times 2$ matrix $W$, a $\wedge_1(W)$ gate can be
simulated by a network of the form}

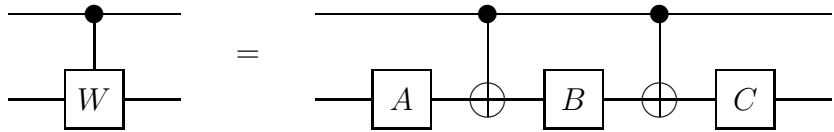
\setlength{\unitlength}{0.030in}

\begin{picture}(40,45)(-10,0)

\put(0,15){\line(1,0){10}}
\put(20,15){\line(1,0){10}}
\put(0,30){\line(1,0){30}}

\put(15,30){\circle*{3}}

\put(15,20){\line(0,1){10}}

\put(10,10){\framebox(10,10){$W$}}

\end{picture}
\begin{picture}(20,45)(0,0)

\put(0,0){\makebox(20,45){$=$}}

\end{picture}
\begin{picture}(90,45)(0,0)

\put(0,15){\line(1,0){10}}
\put(20,15){\line(1,0){20}}
\put(50,15){\line(1,0){20}}
\put(80,15){\line(1,0){10}}
\put(0,30){\line(1,0){90}}

\put(30,15){\circle{6}}
\put(60,15){\circle{6}}

\put(30,30){\circle*{3}}
\put(60,30){\circle*{3}}

\put(30,12){\line(0,1){18}}
\put(60,12){\line(0,1){18}}

\put(10,10){\framebox(10,10){$A$}}
\put(40,10){\framebox(10,10){$B$}}
\put(70,10){\framebox(10,10){$C$}}

\end{picture}

\ni {\sl where $A$, $B$, and $C$ $\in SU(2)$,
if and only if $W\in SU(2)$.}\ee

\loud{Proof:}
For the ``if'' part, let $A$, $B$, and $C$ be as in Lemma 4.3.
If the value of the first (top) bit is 0 then $A \cdot B \cdot C = I$
is applied to the second bit.
If the value of the first bit is 1 then
$A \cdot \sigma_x \cdot B \cdot \sigma_x \cdot C = W$
is applied to the second bit.

For the ``only if'' part, note that $A \cdot B \cdot C = I$ must hold if
the simulation is correct when the first bit is 0.
Also, if the network simulates a $\wedge_1(W)$ gate then
$A \cdot \sigma_x \cdot B \cdot \sigma_x \cdot C = W$.
Therefore, since
${\rm det}(A \cdot\sigma_x\cdot B \cdot\sigma_x\cdot C)=1$,
$W$ must
also be special unitary.\qed\ee

\loud{Lemma 5.2:}{\sl
For any $\d$ and $S = \Scal(\d)$, a $\wedge_1(S)$ gate can be simulated
by a network of the form}

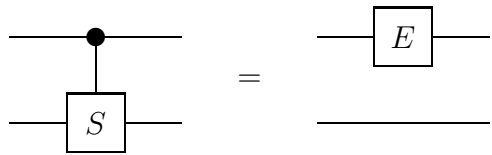
\setlength{\unitlength}{0.030in}

\begin{picture}(60,45)(-30,0)

\put(0,15){\line(1,0){10}}
\put(20,15){\line(1,0){10}}
\put(0,30){\line(1,0){30}}

\put(15,30){\circle*{3}}

\put(15,20){\line(0,1){10}}

\put(10,10){\framebox(10,10){$S$}}

\end{picture}
\begin{picture}(20,45)(0,0)

\put(0,0){\makebox(20,45){$=$}}

\end{picture}
\begin{picture}(30,45)(0,0)

\put(0,15){\line(1,0){30}}
\put(0,30){\line(1,0){10}}
\put(20,30){\line(1,0){10}}

\put(10,25){\framebox(10,10){$E$}}

\end{picture}

\ni {\sl where $E$ is unitary.}\ee

\loud{Proof:}
Let
$$E = \Ph(-\d) \cdot \Scal(\textstyle{\d \over 2}) =
\left(
\begin{array}{cc}
1 & 0 \\
0 & e^{i \d}
\end{array}
\right).$$
Then the observation is that the $4 \times 4$ unitary matrix corresponding
to each of the above networks is
$$\left(
\begin{array}{llll}
1 & 0 & 0 & 0 \\
0 & 1 & 0 & 0 \\
0 & 0 & e^{i \d} & 0 \\
0 & 0 & 0 & e^{i \d}
\end{array}
\right).$$\qed\ee

\ni Clearly, $\wedge_1(S)$ composed with $\wedge_1(W)$ yields
$\wedge_1(S \cdot W)$.
Thus, by noting that any unitary matrix $U$ is of the form $U = S \cdot W$,
where $S = \Scal(\d)$ (for some $\d$) and $W$ is $\in SU(2)$, we obtain
the following.\ee

\loud{Corollary 5.3:}{\sl
For any unitary $2 \times 2$ matrix $U$, a $\wedge_1(U)$ gate can be
simulated by at most six basic gates: four 1-bit gates ($\wedge_0$),
and two XOR gates ($\wedge_1(\sigma_x)$).}\ee

\subsection{Special Cases}

In Section 5.1, we have established a general simulation of a $\wedge_1(U)$
gate for an arbitrary unitary $U$.
For special cases of $U$ that may be of interest, a more efficient
construction than that of Corollary 5.3 is possible.
Clearly, Lemma 5.1 immediately yields a more efficient simulation
for all special unitary matrices.
For example, the ``$x$-axis rotation matrix" (to use the language
suggested by the mapping
between SU(2) and SO(3), the group of rigid-body rotations\cite{Expl})
$${\rm R_x}(\th)=
\left(
\begin{array}{rr}
\cos \th/2 & i \sin \th/2 \\
i \sin \th/2 & \cos \th/2
\end{array}
\right) =
\Ph(\textstyle{\pi \over 2}) \cdot \Rot(\th) \cdot
\Ph(-\textstyle{\pi \over 2})$$
is special unitary.  ($R_x$ is of special interest because $\wedge_2(iR_x)$
is the ``Deutsch gate"\cite{Deu2}, which was shown to be universal
for quantum logic.)
For other specific SU(2) matrices an even more efficient
simulation is possible.\ee

\loud{Lemma 5.4:}{\sl
A $\wedge_1(W)$ gate can be simulated by a network of the form}

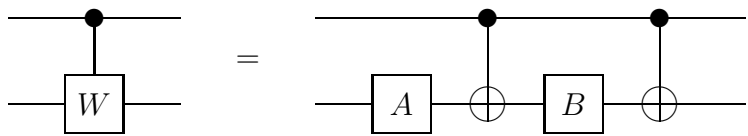
\setlength{\unitlength}{0.030in}

\begin{picture}(40,45)(-10,0)

\put(0,15){\line(1,0){10}}
\put(20,15){\line(1,0){10}}
\put(0,30){\line(1,0){30}}

\put(15,30){\circle*{3}}

\put(15,20){\line(0,1){10}}

\put(10,10){\framebox(10,10){$W$}}

\end{picture}
\begin{picture}(20,45)(0,0)

\put(0,0){\makebox(20,45){$=$}}

\end{picture}
\begin{picture}(75,45)(0,0)

\put(0,15){\line(1,0){10}}
\put(20,15){\line(1,0){20}}
\put(50,15){\line(1,0){25}}
\put(0,30){\line(1,0){75}}

\put(30,15){\circle{6}}
\put(60,15){\circle{6}}

\put(30,30){\circle*{3}}
\put(60,30){\circle*{3}}

\put(30,12){\line(0,1){18}}
\put(60,12){\line(0,1){18}}

\put(10,10){\framebox(10,10){$A$}}
\put(40,10){\framebox(10,10){$B$}}

\end{picture}

\ni {\sl where $A$ and $B$ $\in SU(2)$ if and only if $W$
is of the form}
$$W = \Ph(\a) \cdot \Rot(\th) \cdot
\Ph(\a) =
\left(
\begin{array}{rr}
e^{i \a} \cos \th/2 & \sin \th/2 \\
- \sin \th/2 & e^{-i \a} \cos \th/2
\end{array}
\right),$$
{\sl where $\a$ and $\th$ are real-valued.}\ee

\loud{Proof:}
For the ``if'' part, consider the simulation of $\wedge_1(W)$ that
arises in Lemma 5.1 when
$W = \Ph(\a) \cdot \Rot(\th) \cdot
\Ph(\a)$.
In this case,
$A = \Ph(\a) \cdot \Rot(\textstyle{\th \over 2})$,
$B = \Rot(- \textstyle{\th \over 2}) \cdot \Ph(- \a)$
and $C = I$.
Thus, $B = A^{\scriptsize \dag}$ and $C$ can be omitted.

For the ``only if'' part, note that $B = A^{\scriptsize \dag}$ must hold
for the simulation to be valid when when the first bit is 0.
Therefore, if the first bit is 1 then
$A \cdot \sigma_x \cdot A^{\scriptsize \dag} \cdot \sigma_x$
is applied to the second bit.
Now, the matrix $A \cdot \sigma_x \cdot A^{\scriptsize \dag}$
has determinant $-1$ and is traceless (since its trace is the same as
that of $\sigma_x$).
By specializing the characterization of unitary matrices in Lemma 4.1
to traceless matrices with determinant $-1$, we conclude that
$A \cdot \sigma_x \cdot A^{\scriptsize \dag}$ must be of the form
$$A \cdot \sigma_x \cdot A^{\scriptsize \dag} =
\left(
\begin{array}{rr}
          \sin \th/2 & e^{i \a} \cos \th/2 \\
e^{-i \a} \cos \th/2 &         - \sin \th/2
\end{array}
\right).$$
Therefore,
$$A \cdot \sigma_x \cdot A^{\scriptsize \dag} \cdot \sigma_x =
\left(
\begin{array}{rr}
e^{i \a} \cos \th/2 & \sin \th/2 \\
- \sin \th/2 & e^{-i \a} \cos \th/2
\end{array}
\right),$$
as required.\qed\ee

\ni Examples of matrices of the form of Lemma 5.4 are $\Rot(\th)$ itself,
as well as
$\Ph(\a) =
\Ph(\textstyle{\a \over 2}) \cdot \Rot(0) \cdot \Ph(\textstyle{\a \over 2})$.
However, ${\rm R_x}(\th)$ is not of this form.

Finally, for certain $U$, we obtain an even greater simplification of
the simulation of $\wedge_1(U)$ gates.\ee

\loud{Lemma 5.5:}{\sl
A  $\wedge_1(V)$ gate can be simulated by a construction
of the form}

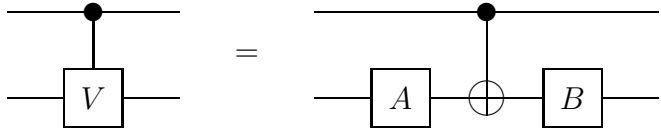
\setlength{\unitlength}{0.030in}

\begin{picture}(40,45)(-10,0)

\put(0,15){\line(1,0){10}}
\put(20,15){\line(1,0){10}}
\put(0,30){\line(1,0){30}}

\put(15,30){\circle*{3}}

\put(15,20){\line(0,1){10}}

\put(10,10){\framebox(10,10){$V$}}

\end{picture}
\begin{picture}(20,45)(0,0)

\put(0,0){\makebox(20,45){$=$}}

\end{picture}
\begin{picture}(60,45)(0,0)

\put(0,15){\line(1,0){10}}
\put(20,15){\line(1,0){20}}
\put(50,15){\line(1,0){10}}
\put(0,30){\line(1,0){60}}

\put(30,15){\circle{6}}

\put(30,30){\circle*{3}}

\put(30,12){\line(0,1){18}}

\put(10,10){\framebox(10,10){$A$}}
\put(40,10){\framebox(10,10){$B$}}

\end{picture}

\ni {\sl where $A$ and $B$ are unitary if and only if $V$ is of the form}
$$V = \Ph(\a) \cdot \Rot(\th) \cdot
\Ph(\a) \cdot \sigma_x
= \left(
\begin{array}{rr}
          \sin \th/2 & e^{i \a} \cos \th/2 \\
e^{-i \a} \cos \th/2 &         - \sin \th/2
\end{array}
\right),$$
{\sl where $\a$ and $\th$ are real-valued.}\ee

\loud{Proof:}
If an additional $\wedge_1(\sigma_x)$ is appended to the end of the network
in Lemma 5.4 then, the network is equivalent to that above (since
$\wedge_1(\sigma_x)$ is an involution), and also simulates a
$\wedge_1(W \cdot \sigma_x)$ gate (since $\wedge_1(W)$ composed with
$\wedge_1(\sigma_x)$ is $\wedge_1(W \cdot \sigma_x)$).\qed\ee

\ni Examples of matrices of the form of Lemma 5.5 are the Pauli matrices
$$\sigma_y =
\left(
\begin{array}{rr}
0 & -i \\
i & 0
\end{array}
\right) =
\Ph(\textstyle{\pi \over 2}) \cdot \Rot(2\pi) \cdot
\Ph(\textstyle{\pi \over 2}) \cdot \sigma_x$$
and
$$\sigma_z =
\left(
\begin{array}{rr}
1 & 0 \\
0 & -1
\end{array}
\right) =
\Ph(0) \cdot \Rot(\pi) \cdot \Ph(0) \cdot \sigma_x$$
(as well as $\sigma_x$ itself).

Lemma 5.5 permits an immediate generalization of Corollary 5.3:\ee

\loud{Corollary 5.6:}{\sl
For any unitary $2 \times 2$ matrix $U$, a $\wedge_1(U)$ gate can be
simulated by at most six basic gates: four 1-bit gates ($\wedge_0$),
and two gates ($\wedge_1(V)$), where $V$ is of the form
$V = \Ph(\a) \cdot \Rot(\th) \cdot
\Ph(\a) \cdot \sigma_x$.}\ee

A particular feature of the $\wedge_1(\sigma_z)$ gates is that they are
symmetric with respect to their input bits.
In view of this, as well as for future reference, we introduce the following
special notation for $\wedge_1(\sigma_z)$ gates.

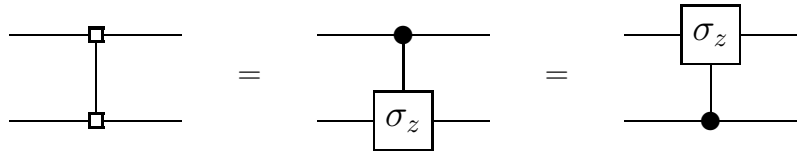
\setlength{\unitlength}{0.030in}

\begin{picture}(45,45)(-15,0)

\put(0,15){\line(1,0){14}}
\put(16,15){\line(1,0){14}}
\put(0,30){\line(1,0){14}}
\put(16,30){\line(1,0){14}}

\put(15,16){\line(0,1){13}}

\thicklines
\put(14,14){\framebox(2,2){}}
\put(14,29){\framebox(2,2){}}
\thinlines

\end{picture}
\begin{picture}(20,45)(0,0)

\put(0,0){\makebox(20,45){$=$}}

\end{picture}
\begin{picture}(30,45)(0,0)

\put(0,15){\line(1,0){10}}
\put(20,15){\line(1,0){10}}
\put(0,30){\line(1,0){30}}

\put(15,20){\line(0,1){10}}

\put(10,10){\framebox(10,10){\large $\sigma_z$}}

\put(15,30){\circle*{3}}

\end{picture}
\begin{picture}(20,45)(0,0)

\put(0,0){\makebox(20,45){$=$}}

\end{picture}
\begin{picture}(30,45)(0,0)

\put(0,15){\line(1,0){30}}
\put(0,30){\line(1,0){10}}
\put(20,30){\line(1,0){10}}

\put(15,15){\line(0,1){10}}

\put(10,25){\framebox(10,10){\large $\sigma_z$}}

\put(15,15){\circle*{3}}

\end{picture}

\section{Three-Bit Networks}

\subsection{Simulation of General $\wedge_2(U)$ Gates}

\loud{Lemma 6.1:}{\sl
For any unitary $2 \times 2$ matrix $U$, a $\wedge_2(U)$ gate can be
simulated by a network of the form}

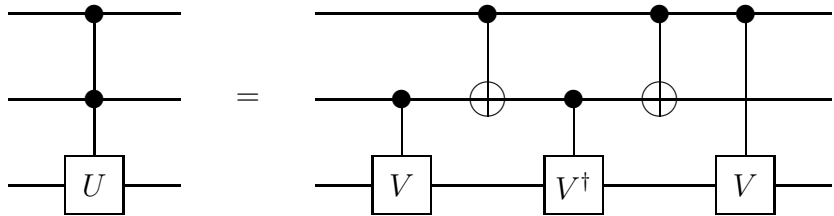
\setlength{\unitlength}{0.030in}

\begin{picture}(40,60)(-10,0)

\put(0,15){\line(1,0){10}}
\put(20,15){\line(1,0){10}}
\put(0,30){\line(1,0){30}}
\put(0,45){\line(1,0){30}}

\put(15,30){\circle*{3}}
\put(15,45){\circle*{3}}

\put(15,20){\line(0,1){25}}

\put(10,10){\framebox(10,10){$U$}}

\end{picture}
\begin{picture}(20,60)(0,0)

\put(0,0){\makebox(20,60){$=$}}

\end{picture}
\begin{picture}(90,60)(0,0)

\put(0,15){\line(1,0){10}}
\put(20,15){\line(1,0){20}}
\put(50,15){\line(1,0){20}}
\put(80,15){\line(1,0){10}}
\put(0,30){\line(1,0){90}}
\put(0,45){\line(1,0){90}}

\put(30,30){\circle{6}}
\put(60,30){\circle{6}}

\put(15,30){\circle*{3}}
\put(45,30){\circle*{3}}
\put(30,45){\circle*{3}}
\put(60,45){\circle*{3}}
\put(75,45){\circle*{3}}

\put(75,20){\line(0,1){25}}
\put(15,20){\line(0,1){10}}
\put(30,27){\line(0,1){18}}
\put(45,20){\line(0,1){10}}
\put(60,27){\line(0,1){18}}

\put(10,10){\framebox(10,10){$V$}}
\put(70,10){\framebox(10,10){$V$}}
\put(40,10){\framebox(10,10){$V^{\scriptsize \dag}$}}

\end{picture}

\ni {\sl where $V$ is unitary.}\ee

\loud{Proof:}
Let $V$ be such that $V^2 = U$.
If the first bit or the second bit are 0 then the transformation
applied to the third bit is either $I$ or $V \cdot V^{\scriptsize \dag} = I$.
If the first two bits are both 1 then the transformation applied to
the third is $V \cdot V = U$.\qed\ee

\ni Some of the intuition behind the construction in the above Lemma is that,
when the first two input bits are $x_1$ and $x_2$, the sequence of operations
performed on the third bit is:
$V$ iff $x_1 = 1$, $V$ iff $x_2 = 1$, and
$V^{\scriptsize \dag}$ iff $x_1 \oplus x_2 = 1$.
Since
$$x_1 + x_2 - (x_1 \oplus x_2) = 2 \cdot (x_1 \wedge x_2)$$
(where ``$+$'', ``$-$'', and ``$\cdot$'' are the ordinary arithmetic
operations), the above sequence of operations is equivalent to performing
$V^2$ on the third bit iff $x_1 \wedge x_2 = 1$, which is the
$\wedge_2(V^2)$ gate.
(This approach generalizes to produce a simulation of $\wedge_m(V^{2^{m-1}})$,
for $m > 2$, which is considered in Section 7.)

We can now combine Lemma 6.1 with Corollary 5.3 to obtain a simulation
of $\wedge_2(U)$ using only basic gates ($\wedge_1(\sigma_x)$ and $\wedge_0$).
The number of these gates is reduced when it is recognized that a number
of the one-bit gates can be {\it merged} and eliminated.  In particular, the
$\wedge_0(C)$ from the end of the simulation of the first $\wedge_1(V)$ gate,
and the $\wedge_0(C^\adj)$ from the $\wedge_1(V^\adj)$ gate combine to
form the identity and are eliminated entirely.  This same sort of merging
occurs to eliminate a $\wedge_0(A)$ gate and a $\wedge_0(A^\adj)$ gate.
We arrive at the following count:\ee

\loud{Corollary 6.2:}{\sl
For any unitary $2 \times 2$ matrix $U$, a $\wedge_2(U)$ gate can be
simulated by at most sixteen basic gates: eight 1-bit gates ($\wedge_0$)
and eight XOR gates ($\wedge_1(\sigma_x)$).}\ee

\ni A noteworthy case is when $U = \sigma_x$, where we obtain a simulation
of the 3-bit Toffoli gate $\wedge_2(\sigma_x)$, which is the primitive gate
for classical reversible logic \cite{Toff}.  Later we will use the fact that
because $\wedge_2(\sigma_x)$ is its own inverse, either the simulation of
Lemma 6.1 or the time-reversed simulation (in which the order of the gates
is reversed, and each unitary operator is replaced by
its Hermitian conjugate) may be used.

\subsection{Three-bit gates congruent to $\wedge_2(U)$}

We now show that more efficient simulations of three-bit gates are
possible if phase shifts of the quantum states other than zero are
permitted.  If we define the matrix $W$ as
$$W = \left(
\begin{array}{rr}
 0 & 1 \\
-1 & 0
\end{array}
\right) = \Scal(\textstyle{\pi \over 2}) \cdot \sigma_y,$$
then the gates $\wedge_2(W)$ and $\wedge_2(\sigma_x)$ can be regarded
as being ``congruent modulo phase shifts'', because the latter gate
differs only in that it maps $|111\ran$ to $-|110\ran$ (instead of
$|110\ran$).
This is perfectly acceptable if the gate is part of an operation which
merely mimics classical reversible computation, or if the gate is paired
with another similar one to cancel out the extra phase, as is sometimes
the case in reversible gate arrangements (see Corollary 7.4);
however, this phase difference
is dangerous in general if non-classical unitary operations appear in the
computation.
Gates congruent to $\wedge_2(\sigma_x)$ modulo phase shifts have been
previously investigated in \cite{Smol}.

The following is a more efficient simulation of a gate congruent
to $\wedge_2(\sigma_x)$ modulo phase shifts:

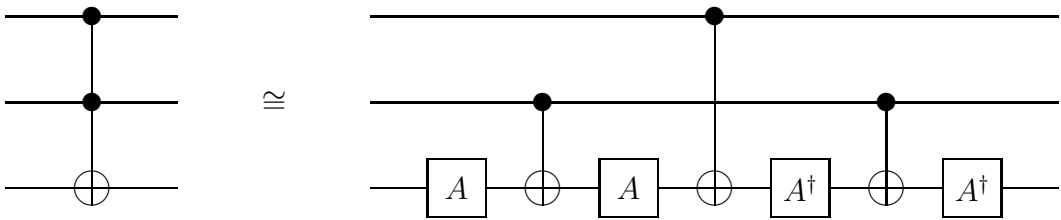
\setlength{\unitlength}{0.030in}

\begin{picture}(30,60)(0,0)

\put(0,15){\line(1,0){30}}
\put(0,30){\line(1,0){30}}
\put(0,45){\line(1,0){30}}

\put(15,12){\line(0,1){33}}

\put(15,15){\circle{6}}

\put(15,30){\circle*{3}}
\put(15,45){\circle*{3}}

\end{picture}
\begin{picture}(30,60)(0,0)

\put(0,0){\makebox(30,60){$\cong$}}

\end{picture}
\begin{picture}(120,45)(0,0)

\put(0,15){\line(1,0){10}}
\put(20,15){\line(1,0){20}}
\put(50,15){\line(1,0){20}}
\put(80,15){\line(1,0){20}}
\put(110,15){\line(1,0){10}}
\put(0,30){\line(1,0){120}}
\put(0,45){\line(1,0){120}}

\put(30,15){\circle{6}}
\put(60,15){\circle{6}}
\put(90,15){\circle{6}}

\put(30,30){\circle*{3}}
\put(60,45){\circle*{3}}
\put(90,30){\circle*{3}}

\put(30,12){\line(0,1){18}}
\put(60,12){\line(0,1){32}}
\put(90,12){\line(0,1){18}}

\put(10,10){\framebox(10,10){$A$}}
\put(40,10){\framebox(10,10){$A$}}
\put(70,10){\framebox(10,10){$A^{\scriptsize \dag}$}}
\put(100,10){\framebox(10,10){$A^{\scriptsize \dag}$}}

\end{picture}

\ni where $A = \Rot(\textstyle{\pi \over 4})$.
In the above, the ``$\cong$'' indicates that the networks are not
identical, but differ at most in the phases of their amplitudes,
which are all $\pm 1$ (the phase of the $|101\rangle$ state
is reversed in this case).

An alternative simulation of a gate congruent to $\wedge_2(\sigma_x)$
modulo phase shifts (whose phase shifts are identical to the previous one)
is given by

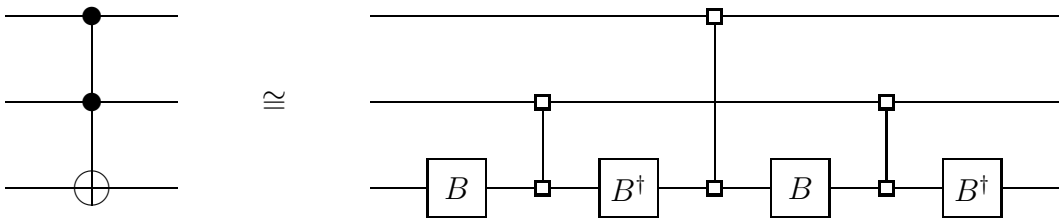
\setlength{\unitlength}{0.030in}

\begin{picture}(30,60)(0,0)

\put(0,15){\line(1,0){30}}
\put(0,30){\line(1,0){30}}
\put(0,45){\line(1,0){30}}

\put(15,12){\line(0,1){33}}

\put(15,15){\circle{6}}

\put(15,30){\circle*{3}}
\put(15,45){\circle*{3}}

\end{picture}
\begin{picture}(30,60)(0,0)

\put(0,0){\makebox(30,60){$\cong$}}

\end{picture}
\begin{picture}(120,60)(0,0)

\put(0,15){\line(1,0){10}}
\put(20,15){\line(1,0){9}}
\put(31,15){\line(1,0){9}}
\put(50,15){\line(1,0){9}}
\put(61,15){\line(1,0){9}}
\put(80,15){\line(1,0){9}}
\put(91,15){\line(1,0){9}}
\put(110,15){\line(1,0){10}}
\put(0,30){\line(1,0){29}}
\put(31,30){\line(1,0){58}}
\put(91,30){\line(1,0){29}}
\put(0,45){\line(1,0){59}}
\put(61,45){\line(1,0){59}}

\put(30,16){\line(0,1){13}}
\thicklines
\put(29,14){\framebox(2,2){}}
\put(29,29){\framebox(2,2){}}
\thinlines
\put(60,16){\line(0,1){28}}
\thicklines
\put(59,14){\framebox(2,2){}}
\put(59,44){\framebox(2,2){}}
\thinlines
\put(90,16){\line(0,1){13}}
\thicklines
\put(89,14){\framebox(2,2){}}
\put(89,29){\framebox(2,2){}}
\thinlines

\put(10,10){\framebox(10,10){$B$}}
\put(40,10){\framebox(10,10){$B^{\scriptsize \dag}$}}
\put(70,10){\framebox(10,10){$B$}}
\put(100,10){\framebox(10,10){$B^{\scriptsize \dag}$}}

\end{picture}

\ni where $B = \Rot(\textstyle{3 \pi \over 4})$.

\section{$n$-Bit Networks}

The technique for simulating $\wedge_2(U)$ gates in Lemma 6.1
generalizes to $\wedge_m(U)$ gates for $m > 2$.
For example, to simulate a $\wedge_3(U)$ gate for any unitary $U$, set
$V$ so that $V^4 = U$ and then construct a network as follows.

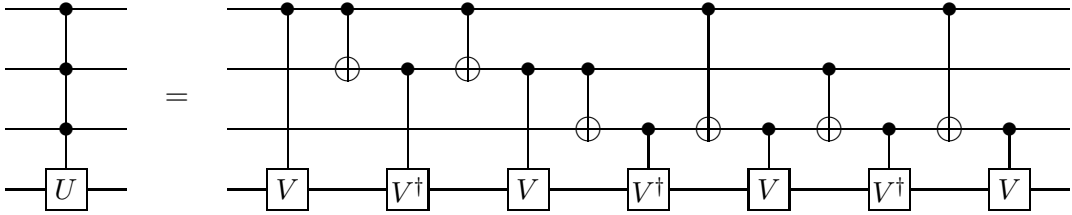
\setlength{\unitlength}{0.021in}

\begin{picture}(30,75)(0,0)

\put(0,15){\line(1,0){10}}
\put(20,15){\line(1,0){10}}
\put(0,30){\line(1,0){30}}
\put(0,45){\line(1,0){30}}
\put(0,60){\line(1,0){30}}

\put(15,20){\line(0,1){40}}

\put(10,10){\small \framebox(10,10){$U$}}

\put(15,30){\circle*{3}}
\put(15,45){\circle*{3}}
\put(15,60){\circle*{3}}

\end{picture}
\begin{picture}(20,75)(0,0)

\put(0,0){\makebox(20,75){$=$}}

\end{picture}
\begin{picture}(210,75)(0,0)

\put(0,15){\line(1,0){10}}
\put(20,15){\line(1,0){20}}
\put(50,15){\line(1,0){20}}
\put(80,15){\line(1,0){20}}
\put(110,15){\line(1,0){20}}
\put(140,15){\line(1,0){20}}
\put(170,15){\line(1,0){20}}
\put(200,15){\line(1,0){10}}
\put(0,30){\line(1,0){210}}
\put(0,45){\line(1,0){210}}
\put(0,60){\line(1,0){210}}

\put(15,60){\circle*{3}}
\put(30,60){\circle*{3}}
\put(45,45){\circle*{3}}
\put(60,60){\circle*{3}}
\put(75,45){\circle*{3}}
\put(90,45){\circle*{3}}
\put(105,30){\circle*{3}}
\put(120,60){\circle*{3}}
\put(135,30){\circle*{3}}
\put(150,45){\circle*{3}}
\put(165,30){\circle*{3}}
\put(180,60){\circle*{3}}
\put(195,30){\circle*{3}}

\put(30,45){\circle{6}}
\put(60,45){\circle{6}}
\put(90,30){\circle{6}}
\put(120,30){\circle{6}}
\put(150,30){\circle{6}}
\put(180,30){\circle{6}}

\put(15,20){\line(0,1){40}}
\put(45,20){\line(0,1){25}}
\put(75,20){\line(0,1){25}}
\put(105,20){\line(0,1){10}}
\put(135,20){\line(0,1){10}}
\put(165,20){\line(0,1){10}}
\put(195,20){\line(0,1){10}}

\put(30,42){\line(0,1){18}}
\put(60,42){\line(0,1){18}}
\put(90,27){\line(0,1){18}}
\put(120,27){\line(0,1){33}}
\put(150,27){\line(0,1){18}}
\put(180,27){\line(0,1){33}}

\put(10,10){\framebox(10,10){\small $V$}}
\put(40,10){\framebox(10,10){\small $V^{\scriptsize \dag}$}}
\put(70,10){\framebox(10,10){\small $V$}}
\put(100,10){\framebox(10,10){\small $V^{\scriptsize \dag}$}}
\put(130,10){\framebox(10,10){\small $V$}}
\put(160,10){\framebox(10,10){\small $V^{\scriptsize \dag}$}}
\put(190,10){\framebox(10,10){\small $V$}}

\end{picture}

\ni The intuition behind this construction is similar to that behind
the construction of Lemma 6.1.
If the first three input bits are $x_1$, $x_2$, and $x_3$ then the
sequence of operations performed on the fourth bit is:

\smallskip
\begin{tabular}{lll}
$V$                    & iff $x_1                       = 1$ & (100) \\
$V^{\scriptsize \dag}$ & iff $x_1 \oplus x_2            = 1$ & (110) \\
$V$                    & iff $x_2                       = 1$ & (010) \\
$V^{\scriptsize \dag}$ & iff $x_2 \oplus x_3            = 1$ & (011) \\
$V$                    & iff $x_1 \oplus x_2 \oplus x_3 = 1$ & (111) \\
$V^{\scriptsize \dag}$ & iff $x_1 \oplus x_3            = 1$ & (101) \\
$V$                    & iff $x_3                       = 1$ & (001).
\end{tabular}

\smallskip
\ni The strings on the right encode the condition for the operation $V$
or $V^\adj$ at each step---the ``1"'s indicate which input
bits are involved in
the condition.  For an efficient implementation of $\wedge_3(U)$, these
strings form a grey code sequence.  Note also that the parity of each
bit string
determines whether to apply $V$ or $V^{\scriptsize \dag}$.
By comparing this sequence
of operations with the terms in the equation
$$x_1 + x_2 + x_3 - (x_1 \oplus x_2) - (x_1 \oplus x_3) - (x_2 \oplus x_3)
+ (x_1 \oplus x_2 \oplus x_3) = 4 \cdot (x_1 \wedge x_2 \wedge x_3),$$
it can be verified that the above sequence of operations is equivalent to
performing $V^4$ on the fourth bit iff $x_1 \wedge x_2 \wedge x_3 = 1$,
which is the $\wedge_3(V^4)$ gate.

The foregoing can be generalized to simulate $\wedge_m(U)$ for
larger values of $m$.\ee

\loud{Lemma 7.1:}{\sl
For any $n \ge 3$ and any unitary $2 \times 2$ matrix $U$,
a $\wedge_{n-1}(U)$ gate can be simulated by an $n$-bit network
consisting of $2^{n-1} - 1$ $\wedge_1(V)$ and
$\wedge_1(V^{\scriptsize \dag})$ gates and
$2^{n-1} - 2$ $\wedge_1(\sigma_x)$ gates, where $V$ is unitary.}\ee

\ni We omit the proof of Lemma 7.1, but point out that it is a generalization
of the $n=4$ case above and based on setting $V$ so that $V^{2^{n-2}} = U$
and
``implementing" the identity
\begin{eqnarray*}
\sum_{k_1} x_{k_1}
- \sum_{k_1 < k_2} (x_{k_1} \oplus x_{k_2})
+ \sum_{k_1 < k_2 < k_3} (x_{k_1} \oplus x_{k_2} \oplus x_{k_3})
- \cdots + (-1)^{m-1} (x_1 \oplus x_2 \oplus \cdots \oplus x_m) & & \\
= 2^{m-1} \cdot (x_1 \wedge x_2 \wedge \cdots \wedge x_m)
\mbox{\ \ \ \ \ \ \ \ \ \ \ \ \ \ \ \ \ \ \ \ \ \ \ \ \ \ \ \ \ \ \ \ \
\ \ \ \ \ \ \ \ \ \ \ \ \ \ \ \ \ \ \ \ \ \ \ \ \ \ \ \ \ \ \ \ \ \ \ \
\ \ \ \ \ \ } & &
\end{eqnarray*}
with a grey-code sequence of operations.

For some specific small values of $n$
(for $n = \mbox{3, 4, 5, 6, 7, and 8}$), this is the most efficient
technique that we are aware of for simulating arbitrary $\wedge_{n-1}(U)$
gates as well as $\wedge_{n-1}(\sigma_x)$ gates;
taking account of mergers (see Corollary 6.2), the simulation
requires $3\cdot 2^{n-1}-4$ $\wedge_1(\sigma_x)$'s and
$2\cdot 2^{n-1}$ $\wedge_0$'s.  However, since this number is
$\Theta(2^n)$, the simulation is very inefficient for large values of $n$.
For the remainder of this section, we focus on the asymptotic
growth rate of the simulations with respect to $n$, and show that this
can be quadratic in the general case and linear in many cases of interest.

\subsection{Linear Simulation of $\wedge_{n-2}(\sigma_x)$ Gates on $n$-Bit
Networks}

\loud{Lemma 7.2:}{\sl
If $n \ge 5$ and $m \in \{3,\ldots,\lceil {n \over 2} \rceil \}$ then
a $\wedge_m(\sigma_x)$ gate can be simulated by a network consisting of
$4(m-2)$ $\wedge_2(\sigma_x)$ gates that is of the form}

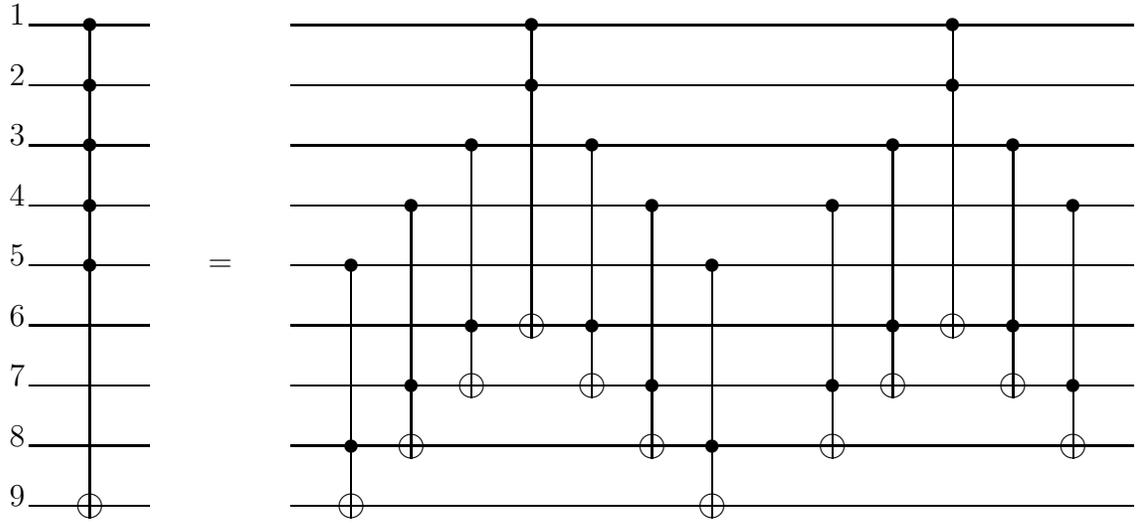
\setlength{\unitlength}{0.021in}

\ni \begin{picture}(40,135)(-10,0)

\put(0,15){\line(1,0){30}}
\put(0,30){\line(1,0){30}}
\put(0,45){\line(1,0){30}}
\put(0,60){\line(1,0){30}}
\put(0,75){\line(1,0){30}}
\put(0,90){\line(1,0){30}}
\put(0,105){\line(1,0){30}}
\put(0,120){\line(1,0){30}}
\put(0,135){\line(1,0){30}}

\put(-5,15){9}
\put(-5,30){8}
\put(-5,45){7}
\put(-5,60){6}
\put(-5,75){5}
\put(-5,90){4}
\put(-5,105){3}
\put(-5,120){2}
\put(-5,135){1}

\put(15,15){\circle{6}}

\put(15,75){\circle*{3}}
\put(15,90){\circle*{3}}
\put(15,105){\circle*{3}}
\put(15,120){\circle*{3}}
\put(15,135){\circle*{3}}

\put(15,12){\line(0,1){123}}

\end{picture}
\begin{picture}(30,150)(0,0)

\put(0,0){\makebox(30,150){$=$}}

\end{picture}
\begin{picture}(210,150)(0,0)

\put(0,15){\line(1,0){210}}
\put(0,30){\line(1,0){210}}
\put(0,45){\line(1,0){210}}
\put(0,60){\line(1,0){210}}
\put(0,75){\line(1,0){210}}
\put(0,90){\line(1,0){210}}
\put(0,105){\line(1,0){210}}
\put(0,120){\line(1,0){210}}
\put(0,135){\line(1,0){210}}

\put(15,15){\circle{6}}
\put(105,15){\circle{6}}
\put(30,30){\circle{6}}
\put(90,30){\circle{6}}
\put(135,30){\circle{6}}
\put(195,30){\circle{6}}
\put(45,45){\circle{6}}
\put(75,45){\circle{6}}
\put(150,45){\circle{6}}
\put(180,45){\circle{6}}
\put(60,60){\circle{6}}
\put(165,60){\circle{6}}

\put(15,30){\circle*{3}}
\put(105,30){\circle*{3}}
\put(30,45){\circle*{3}}
\put(90,45){\circle*{3}}
\put(135,45){\circle*{3}}
\put(195,45){\circle*{3}}
\put(45,60){\circle*{3}}
\put(75,60){\circle*{3}}
\put(150,60){\circle*{3}}
\put(180,60){\circle*{3}}
\put(15,75){\circle*{3}}
\put(105,75){\circle*{3}}
\put(30,90){\circle*{3}}
\put(90,90){\circle*{3}}
\put(135,90){\circle*{3}}
\put(195,90){\circle*{3}}
\put(45,105){\circle*{3}}
\put(75,105){\circle*{3}}
\put(150,105){\circle*{3}}
\put(180,105){\circle*{3}}
\put(60,120){\circle*{3}}
\put(60,135){\circle*{3}}
\put(165,120){\circle*{3}}
\put(165,135){\circle*{3}}

\put(15,12){\line(0,1){63}}
\put(30,27){\line(0,1){63}}
\put(45,42){\line(0,1){63}}
\put(60,57){\line(0,1){78}}
\put(75,42){\line(0,1){63}}
\put(90,27){\line(0,1){63}}
\put(105,12){\line(0,1){63}}
\put(135,27){\line(0,1){63}}
\put(150,42){\line(0,1){63}}
\put(165,57){\line(0,1){78}}
\put(180,42){\line(0,1){63}}
\put(195,27){\line(0,1){63}}

\end{picture}

\ni {\sl (illustrated for $n = 9$ and $m = 5$).}\ee

\loud{Proof:}
Consider the group of the first 7 gates in the above network.
The sixth bit (from the top) is negated iff the first two bits are 1,
the seventh bit is negated iff the first three bits are 1, the eighth
bit is negated iff the first four bits are 1, and the ninth bit is negated
iff the first five bits are 1.
Thus, the last bit is correctly set, but the three preceding bits are
altered.
The last 5 gates in the network reset the values of these three preceding
bits.\qed\ee

\ni Note that in this construction and in the ones following, although
many of the bits not involved in the gate are operated upon, the gate
operation is performed  correctly independent of the initial state
of the bits (i.e., they do not have to be ``cleared" to 0 first), and
they are reset to their
initial values after the operations of the gate
(as in the computations which occur in \cite{BenO} and \cite{Clev}).
This fact makes constructions like the following possible.\ee

\loud{Lemma 7.3:}{\sl
For any $n \ge 5$, and $m \in \{2,\ldots,n-3\}$ a $\wedge_{n-2}(\sigma_x)$
gate can be simulated by a network consisting of two $\wedge_m(\sigma_x)$
gates and two $\wedge_{n-m-1}(\sigma_x)$ gates which is of the form}

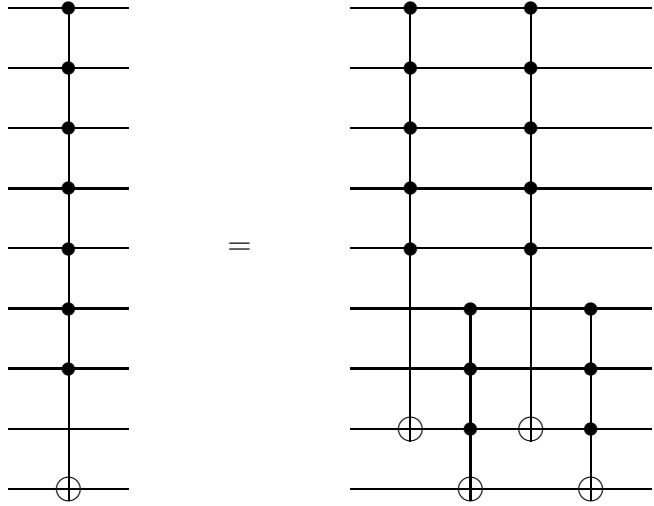
\begin{picture}(80,150)(-50,0)

\put(0,15){\line(1,0){30}}
\put(0,30){\line(1,0){30}}
\put(0,45){\line(1,0){30}}
\put(0,60){\line(1,0){30}}
\put(0,75){\line(1,0){30}}
\put(0,90){\line(1,0){30}}
\put(0,105){\line(1,0){30}}
\put(0,120){\line(1,0){30}}
\put(0,135){\line(1,0){30}}

\put(15,15){\circle{6}}

\put(15,45){\circle*{3}}
\put(15,60){\circle*{3}}
\put(15,75){\circle*{3}}
\put(15,90){\circle*{3}}
\put(15,105){\circle*{3}}
\put(15,120){\circle*{3}}
\put(15,135){\circle*{3}}

\put(15,12){\line(0,1){123}}

\end{picture}
\begin{picture}(50,150)(0,0)

\put(0,0){\makebox(50,150){$=$}}

\end{picture}
\begin{picture}(75,150)(0,0)

\put(0,15){\line(1,0){75}}
\put(0,30){\line(1,0){75}}
\put(0,45){\line(1,0){75}}
\put(0,60){\line(1,0){75}}
\put(0,75){\line(1,0){75}}
\put(0,90){\line(1,0){75}}
\put(0,105){\line(1,0){75}}
\put(0,120){\line(1,0){75}}
\put(0,135){\line(1,0){75}}

\put(30,15){\circle{6}}
\put(60,15){\circle{6}}
\put(15,30){\circle{6}}
\put(45,30){\circle{6}}

\put(15,75){\circle*{3}}
\put(15,90){\circle*{3}}
\put(15,105){\circle*{3}}
\put(15,120){\circle*{3}}
\put(15,135){\circle*{3}}

\put(30,30){\circle*{3}}
\put(30,45){\circle*{3}}
\put(30,60){\circle*{3}}

\put(45,75){\circle*{3}}
\put(45,90){\circle*{3}}
\put(45,105){\circle*{3}}
\put(45,120){\circle*{3}}
\put(45,135){\circle*{3}}

\put(60,30){\circle*{3}}
\put(60,45){\circle*{3}}
\put(60,60){\circle*{3}}

\put(15,27){\line(0,1){108}}
\put(30,12){\line(0,1){48}}
\put(45,27){\line(0,1){108}}
\put(60,12){\line(0,1){48}}

\end{picture}

\ni {\sl (illustrated for $n = 9$ and $m = 5$).}\ee

\loud{Proof:}
By inspection.\qed\ee

\loud{Corollary 7.4:}{\sl
On an $n$-bit network (where $n \ge 7$), a $\wedge_{n-2}(\sigma_x)$ gate
can be simulated by $8(n-5)$ $\wedge_2(\sigma_x)$ gates (3-bit Toffoli
gates), as well as by $48n-204$ basic operations.}\ee

\loud{Proof:}
First apply Lemma 7.2 with $m_1 = \lceil {n \over 2} \rceil$ and
$m_2 = n - m_1 - 1$ to simulate $\wedge_{m_1}(\sigma_x)$ and
$\wedge_{m_2}(\sigma_x)$ gates.
Then combine these by Lemma 7.3 to simulate the $\wedge_{n-2}(\sigma_x)$
gate.
Then, each $\wedge_2(\sigma_x)$ gate in the above simulation may be
simulated by a set of basic operations (as in Corollary 6.2).  We find that
almost all of these Toffoli gates need only to be simulated modulo phase
factors as in Sec. 6.2; in particular, only 4 of the Toffoli gates, the
ones which involve the last bit in the diagram above, need to be simulated
exactly according to the construction of Corollary 6.2.  Thus these 4 gates
are simulated by 16 basic operations, while the other $8n-36$ Toffoli gates
are simulated in just 6 basic operations.  A careful accounting of the
mergers of $\wedge_0$ gates which are then possible leads to the total
count of basic operations given above.
\qed\ee

\ni The above constructions, though asymptotically efficient, requires at
least one ``extra'' bit, in that an $n$-bit network is required
to simulate the $(n-1)$-bit gate $\wedge_{n-2}(\sigma_x)$.
In the next subsection, we shall show how to construct $\wedge_{n-1}(U)$
for an arbitrary unitary $U$ using a quadratic number of basic operations
on an $n$-bit network, which includes the $n$-bit Toffoli gate
$\wedge_{n-1}(\sigma_x)$ as a special case.

\subsection{Quadratic Simulation of General $\wedge_{n-1}(U)$ Gates on
$n$-Bit Networks}

\loud{Lemma 7.5:}{\sl
For any unitary $2 \times 2$ matrix $U$, a $\wedge_{n-1}(U)$ gate can
be simulated by a network of the form}

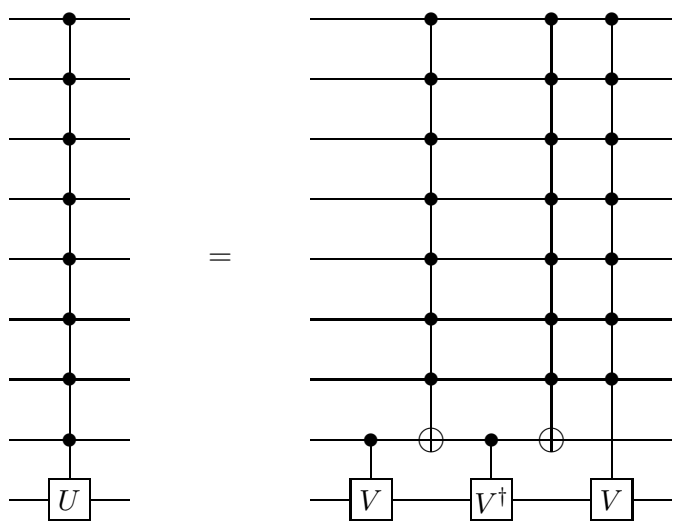
\setlength{\unitlength}{0.021in}

\begin{picture}(60,150)(-30,0)

\put(0,15){\line(1,0){10}}
\put(20,15){\line(1,0){10}}
\put(0,30){\line(1,0){30}}
\put(0,45){\line(1,0){30}}
\put(0,60){\line(1,0){30}}
\put(0,75){\line(1,0){30}}
\put(0,90){\line(1,0){30}}
\put(0,105){\line(1,0){30}}
\put(0,120){\line(1,0){30}}
\put(0,135){\line(1,0){30}}

\put(15,30){\circle*{3}}
\put(15,45){\circle*{3}}
\put(15,60){\circle*{3}}
\put(15,75){\circle*{3}}
\put(15,90){\circle*{3}}
\put(15,105){\circle*{3}}
\put(15,120){\circle*{3}}
\put(15,135){\circle*{3}}

\put(15,20){\line(0,1){115}}

\put(10,10){\framebox(10,10){\small $U$}}

\end{picture}
\begin{picture}(40,150)(0,0)

\put(0,0){\makebox(40,150){$=$}}

\end{picture}
\begin{picture}(90,150)(0,0)

\put(0,15){\line(1,0){10}}
\put(20,15){\line(1,0){20}}
\put(50,15){\line(1,0){20}}
\put(80,15){\line(1,0){10}}
\put(0,30){\line(1,0){90}}
\put(0,45){\line(1,0){90}}
\put(0,60){\line(1,0){90}}
\put(0,75){\line(1,0){90}}
\put(0,90){\line(1,0){90}}
\put(0,105){\line(1,0){90}}
\put(0,120){\line(1,0){90}}
\put(0,135){\line(1,0){90}}

\put(30,30){\circle{6}}
\put(60,30){\circle{6}}

\put(15,30){\circle*{3}}
\put(45,30){\circle*{3}}
\put(75,45){\circle*{3}}
\put(30,45){\circle*{3}}
\put(60,45){\circle*{3}}
\put(75,60){\circle*{3}}
\put(30,60){\circle*{3}}
\put(60,60){\circle*{3}}
\put(75,75){\circle*{3}}
\put(30,75){\circle*{3}}
\put(60,75){\circle*{3}}
\put(75,90){\circle*{3}}
\put(30,90){\circle*{3}}
\put(60,90){\circle*{3}}
\put(75,105){\circle*{3}}
\put(30,105){\circle*{3}}
\put(60,105){\circle*{3}}
\put(75,120){\circle*{3}}
\put(30,120){\circle*{3}}
\put(60,120){\circle*{3}}
\put(75,135){\circle*{3}}
\put(30,135){\circle*{3}}
\put(60,135){\circle*{3}}

\put(75,20){\line(0,1){115}}
\put(15,20){\line(0,1){10}}
\put(30,27){\line(0,1){108}}
\put(45,20){\line(0,1){10}}
\put(60,27){\line(0,1){108}}

\put(10,10){\framebox(10,10){\small $V$}}
\put(70,10){\framebox(10,10){\small $V$}}
\put(40,10){\framebox(10,10){\small $V^{\scriptsize \dag}$}}

\end{picture}\ee

\ni {\sl (illustrated for $n = 9$), where $V$ is unitary.}\ee

\loud{Proof:}
The proof is very similar to that of Lemma 6.1,
setting $V$ so that $V^2 = U$.\qed\ee

\loud{Corollary 7.6:}{\sl
For any unitary $U$, a $\wedge_{n-1}(U)$ gate can be simulated in terms
of $\Theta(n^2)$ basic operations.}\ee

\loud{Proof:}
This is a recursive application of Lemma 7.5.
Let $C_{n-1}$ denote the cost of simulating a $\wedge_{n-1}(U)$
(for an arbitrary $U$).
Consider the simulation in Lemma 7.5.
The cost of simulating the $\wedge_1(V)$ and $\wedge_1(V^{\scriptsize \dag})$
gates is $\Theta(1)$ (by Corollary 5.3).
The cost of simulating the two $\wedge_{n-2}(\sigma_x)$ gates is
$\Theta(n)$ (by Corollary 7.4).
The cost of simulating the $\wedge_{n-2}(V)$ gate (by a recursive application
of Lemma 7.5) is $C_{n-2}$.
Therefore, $C_{n-1}$ satisfies a recurrence of the form
$$C_{n-1} = C_{n-2} + \Theta(n),$$
which implies that $C_{n-1} \in \Theta(n^2)$.\qed\ee

\ni In fact, we find that using the gate-counting mentioned in Corollary 7.4,
the number of basic operations is $48n^2+O(n)$.

\ni Although Corollary 7.6 is significant in that it permits any
$\wedge_{n-1}(U)$ to be simulated with ``polynomial complexity'',
the question remains as to whether a subquadratic simulation is possible.
The following is an $\Omega(n)$ lower bound on this complexity.\ee

\loud{Lemma 7.7:}{\sl
Any simulation of a nonscalar $\wedge_{n-1}(U)$ gate
(i.e. where $U \not= \Scal(\delta)\cdot I$)
requires at least $n-1$ basic operations.}\ee

\loud{Proof:}
Consider any $n$-bit network with arbitrarily many 1-bit gates and fewer
than $n-1$ $\wedge_1(\sigma_x)$ gates.
Call two bits {\it adjacent\/} if there there is a $\wedge_1(\sigma_x)$
gate between them, and {\it connected\/} if there is a sequence of
consecutively adjacent bits between them.
Since there are fewer than $n-1$ $\wedge_1(\sigma_x)$ gates, it must be
possible to partition the bits into two nonempty sets ${\cal A}$ and
${\cal B}$
such that no bit in ${\cal A}$ is connected to any bit in ${\cal B}$.
This implies that the unitary transformation associated with the network
is of the form $A \otimes B$, where $A$ is
$2^{|{\cal A}|}$-dimensional and $B$ is
$2^{|{\cal B}|}$-dimensional.
Since the transformation $\wedge_{n-1}(U)$ is not of this form, the
network cannot compute $\wedge_{n-1}(U)$.\qed\ee

It is conceivable that a linear size simulation of $\wedge_{n-1}(U)$
gates is possible.
Although we cannot show this presently, in the remaining subsections,
we show that something ``similar'' (in a number of different senses)
to a linear size simulation of $\wedge_{n-1}(U)$ gates is possible.

\subsection{Linear Approximate Simulation of General $\wedge_{n-1}(U)$
Gates on $n$-Bit Networks}

\loud{Definition:}
We say that one network {\it approximates} another one {\it within} $\e$
if the distance (induced by the Euclidean vector norm) between the unitary
transformations associated with the two networks is at most $\e$.\ee

\ni
This notion of approximation in the context of reducing the complexity
of quantum computations was introduced by Coppersmith\cite{Copp}, and is
useful for the following reason.
Suppose that two networks that are approximately the same (in the above
sense) are executed with identical inputs and their outputs are observed.
Then the probability distributions of the two outcomes will be approximately
the same in the sense that, for any event, its probability will differ
by at most $2\epsilon$ between the two networks.\ee

\loud{Lemma 7.8:}{\sl
For any unitary $2 \times 2$ matrix $U$ and $\e > 0$, a $\wedge_{n-1}(U)$
gate can be approximated within $\e$ by $\Theta(n \log({1 \over \e}))$
basic operations.}\ee

\loud{Proof:}
The idea is to apply Lemma 7.5 recursively as in Corollary 7.6, but
to observe that, with suitable choices for $V$, the recurrence can be
terminated after $\Theta(\log({1 \over \e}))$ levels.

Since $U$ is unitary, there exist unitary matrices $P$ and $D$, such that
$U = P^{\scriptsize \dag} \cdot D \cdot P$ and
$$D =
\left(
\begin{array}{ll}
e^{i d_1} & 0         \\
0         & e^{i d_2}
\end{array}
\right)$$
where $d_1$ and $d_2$ are real.
$e^{i d_1}$ and $e^{i d_2}$ are the eigenvalues of $U$.
If $V_k$ is the matrix used in the $k^{\scriptsize \mbox{th}}$ recursive
application of Lemma 7.3 ($k \in \{0,1,2,\dots\}$) then it is sufficient
that $V_{k+1}^2 = V_k$ for each $k \in \{0,1,2,\dots\}$.
Thus, it suffices to set $V_k = P^{\scriptsize \dag} \cdot D_k \cdot P$,
where
$$D_k =
\left(
\begin{array}{ll}
e^{i d_1 / 2^k} & 0               \\
0               & e^{i d_2 / 2^k}
\end{array}
\right),$$
for each $k \in \{0,1,2,\dots\}$.
Note that then
\begin{eqnarray*}
\|V_k - I\|_2 & =   & \|P^{\scriptsize \dag} \cdot D_k \cdot P - I\|_2 \\
              & =   & \|P^{\scriptsize \dag} \cdot (D_k - I) \cdot P\|_2 \\
              & \le & \|P^{\scriptsize \dag}\|_2 \cdot
                      \|D_k - I\|_2 \cdot \|P\|_2 \\
              & =   & \|D_k - I\|_2 \\
              & \le & \textstyle{\pi / 2^k}.
\end{eqnarray*}
Therefore, if the recursion is terminated after
$k = \lceil \log_2({\pi \over \e}) \rceil$ steps then the discrepancy
between what the resulting network computes and $\wedge_{n-1}(U)$ is an
$(n-k)$-bit transformation of the form $\wedge_{n-k-1}(V_k)$.
Since
$\|\wedge_{n-k-1}(V_k) - \wedge_{n-k-1}(I)\|_2 = \|V_k - I\|_2
\le \pi / 2^{\lceil \log_2({\pi \over \e}) \rceil} \le \e$,
the network approximates $\wedge_{n-1}(U)$ within $\e$.\qed\ee

\subsection{Linear Simulation in Special Cases}

\loud{Lemma 7.9:}{\sl
For any SU(2) matrix $W$, a $\wedge_{n-1}(W)$ gate
can be simulated by a network of the form}

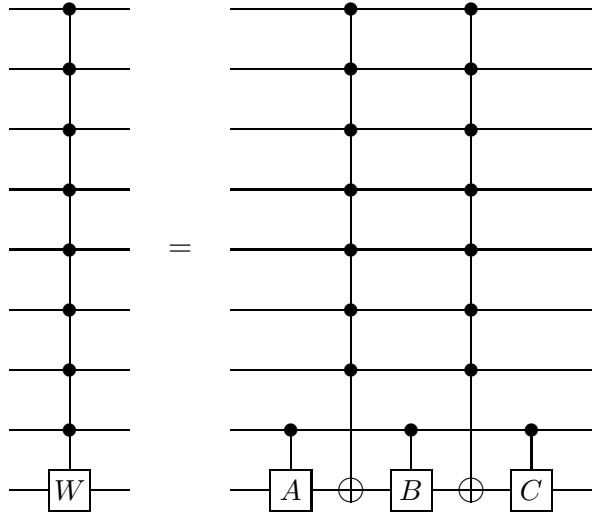
\setlength{\unitlength}{0.021in}

\begin{picture}(60,150)(-30,0)

\put(0,15){\line(1,0){10}}
\put(20,15){\line(1,0){10}}
\put(0,30){\line(1,0){30}}
\put(0,45){\line(1,0){30}}
\put(0,60){\line(1,0){30}}
\put(0,75){\line(1,0){30}}
\put(0,90){\line(1,0){30}}
\put(0,105){\line(1,0){30}}
\put(0,120){\line(1,0){30}}
\put(0,135){\line(1,0){30}}

\put(15,30){\circle*{3}}
\put(15,45){\circle*{3}}
\put(15,60){\circle*{3}}
\put(15,75){\circle*{3}}
\put(15,90){\circle*{3}}
\put(15,105){\circle*{3}}
\put(15,120){\circle*{3}}
\put(15,135){\circle*{3}}

\put(15,20){\line(0,1){115}}

\put(10,10){\framebox(10,10){\small $W$}}

\end{picture}
\begin{picture}(20,150)(0,0)

\put(0,0){\makebox(20,150){$=$}}

\end{picture}
\begin{picture}(90,150)(0,0)

\put(0,15){\line(1,0){10}}
\put(20,15){\line(1,0){20}}
\put(50,15){\line(1,0){20}}
\put(80,15){\line(1,0){10}}
\put(0,30){\line(1,0){90}}
\put(0,45){\line(1,0){90}}
\put(0,60){\line(1,0){90}}
\put(0,75){\line(1,0){90}}
\put(0,90){\line(1,0){90}}
\put(0,105){\line(1,0){90}}
\put(0,120){\line(1,0){90}}
\put(0,135){\line(1,0){90}}

\put(30,15){\circle{6}}
\put(60,15){\circle{6}}

\put(15,30){\circle*{3}}
\put(45,30){\circle*{3}}
\put(75,30){\circle*{3}}
\put(30,45){\circle*{3}}
\put(60,45){\circle*{3}}
\put(30,60){\circle*{3}}
\put(60,60){\circle*{3}}
\put(30,75){\circle*{3}}
\put(60,75){\circle*{3}}
\put(30,90){\circle*{3}}
\put(60,90){\circle*{3}}
\put(30,105){\circle*{3}}
\put(60,105){\circle*{3}}
\put(30,120){\circle*{3}}
\put(60,120){\circle*{3}}
\put(30,135){\circle*{3}}
\put(60,135){\circle*{3}}

\put(30,12){\line(0,1){123}}
\put(60,12){\line(0,1){123}}
\put(15,20){\line(0,1){10}}
\put(45,20){\line(0,1){10}}
\put(75,20){\line(0,1){10}}

\put(10,10){\framebox(10,10){\small $A$}}
\put(40,10){\framebox(10,10){\small $B$}}
\put(70,10){\framebox(10,10){\small $C$}}

\end{picture}

\ni {\sl where $A$, $B$, and $C$ $\in SU(2)$.}\ee

\loud{Proof:}
The proof is very similar to that of Lemma 5.1, referring to
Lemma 4.3.\qed\ee

\ni Combining Lemma 7.9 with Corollary 7.4, we obtain the following.\ee

\loud{Corollary 7.10:}{\sl
For any $W\in SU(2)$, a $\wedge_{n-2}(W)$ gate can be simulated by
$\Theta(n)$ basic operations.}\ee

\ni As in Section 5, a noteworthy example is when
$$W = \left(
\begin{array}{rr}
 0 & 1 \\
-1 & 0
\end{array}
\right) = \Scal(\textstyle{\pi \over 2}) \cdot \sigma_y.$$
In this case, we obtain a linear simulation of a transformation
congruent modulo phase shifts to the $n$-bit Toffoli gate
$\wedge_{n-1}(\sigma_x)$.

\subsection{Linear Simulation of General $\wedge_{n-2}(U)$ Gates on $n$-Bit
Networks With One Bit Fixed}

\loud{Lemma 7.11:}{\sl
For any unitary $U$, a $\wedge_{n-2}(U)$ gate can be simulated by an $n$-bit
network of the form}

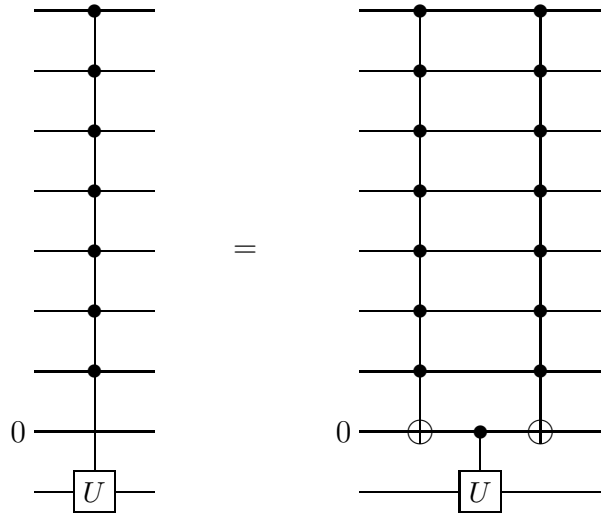
\setlength{\unitlength}{0.021in}

\begin{picture}(60,150)(-30,0)

\put(-6,27.4){0}

\put(0,15){\line(1,0){10}}
\put(20,15){\line(1,0){10}}
\put(0,30){\line(1,0){30}}
\put(0,45){\line(1,0){30}}
\put(0,60){\line(1,0){30}}
\put(0,75){\line(1,0){30}}
\put(0,90){\line(1,0){30}}
\put(0,105){\line(1,0){30}}
\put(0,120){\line(1,0){30}}
\put(0,135){\line(1,0){30}}

\put(15,45){\circle*{3}}
\put(15,60){\circle*{3}}
\put(15,75){\circle*{3}}
\put(15,90){\circle*{3}}
\put(15,105){\circle*{3}}
\put(15,120){\circle*{3}}
\put(15,135){\circle*{3}}

\put(15,20){\line(0,1){115}}

\put(10,10){\framebox(10,10){\small $U$}}

\end{picture}
\begin{picture}(40,150)(0,0)

\put(0,0){\makebox(40,150){$=$}}

\end{picture}
\begin{picture}(66,150)(-6,0)

\put(-6,27.4){0}

\put(0,15){\line(1,0){25}}
\put(35,15){\line(1,0){25}}
\put(0,30){\line(1,0){60}}
\put(0,45){\line(1,0){60}}
\put(0,60){\line(1,0){60}}
\put(0,75){\line(1,0){60}}
\put(0,90){\line(1,0){60}}
\put(0,105){\line(1,0){60}}
\put(0,120){\line(1,0){60}}
\put(0,135){\line(1,0){60}}

\put(15,30){\circle{6}}
\put(45,30){\circle{6}}

\put(30,30){\circle*{3}}
\put(15,45){\circle*{3}}
\put(45,45){\circle*{3}}
\put(15,60){\circle*{3}}
\put(45,60){\circle*{3}}
\put(15,75){\circle*{3}}
\put(45,75){\circle*{3}}
\put(15,90){\circle*{3}}
\put(45,90){\circle*{3}}
\put(15,105){\circle*{3}}
\put(45,105){\circle*{3}}
\put(15,120){\circle*{3}}
\put(45,120){\circle*{3}}
\put(15,135){\circle*{3}}
\put(45,135){\circle*{3}}

\put(15,27){\line(0,1){108}}
\put(30,20){\line(0,1){10}}
\put(45,27){\line(0,1){108}}

\put(25,10){\framebox(10,10){\small $U$}}

\end{picture}

\ni {\sl (illustrated for $n=9$), where the initial value of one bit
(the second to last) is fixed at 0 (and it incurs no net change).}\ee

\loud{Proof:}
By inspection.\qed\ee

\ni Combining Lemma 7.11 with Corollary 7.4, we obtain the following.\ee

\loud{Corollary 7.12:}{\sl
For any unitary $U$, a $\wedge_{n-2}(U)$ gate can be simulated by
$\Theta(n)$ basic operations in $n$-bit network, where the initial
value of one bit is fixed and incurs no net change.}\ee

\ni Note that the ``extra'' bit above may be reused in the course of several
simulations of $\wedge_m(U)$ gates.

\section{Efficient general gate constructions}

In this final discussion we will change the ground rules slightly by
considering the
``basic operation" to be {\it any} two-bit operation.  This may or may
not be a physically reasonable choice in various particular implementations
of quantum computing, but for the moment this should be considered as
just a mathematical convenience which will permit us to address somewhat
more
general questions than the ones considered above.  When the arbitrary
two-bit
gate is taken as the basic operation, then as we have seen, 5 operations
suffice to produce the Toffoli gate (recall Lemma 6.1), 3 produce
the Toffoli gate modulo phases (we permit a merging of the operations in
the construction of Sec. 6.2), and 13 can be used to produce the 4-bit
Toffoli
gate (see Lemma 7.1).
In no case do we have a proof that this is the most
economical method for producing each of these functions; however, for most
of these examples we
have compelling evidence from numerical study that these are in fact
minimal\cite{Smol}.

In the course of doing these numerical investigations we discovered a
number
of interesting additional facts about two-bit gate constructions.  It is
natural to ask, how many two-bit gates are required to perform {\it any
arbitrary} three-bit unitary operation, if the two-bit gates were permitted
to implement any member of U(4)?  The answer is six, as in the gate
arrangement shown here.

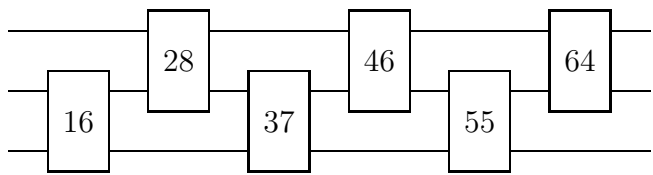
\begin{picture}(160,60)(-20,0)

\put(0,15){\line(1,0){10}}
\put(25,15){\line(1,0){35}}
\put(75,15){\line(1,0){35}}
\put(125,15){\line(1,0){35}}
\put(0,30){\line(1,0){10}}
\put(25,30){\line(1,0){10}}
\put(50,30){\line(1,0){10}}
\put(75,30){\line(1,0){10}}
\put(100,30){\line(1,0){10}}
\put(125,30){\line(1,0){10}}
\put(150,30){\line(1,0){10}}
\put(0,45){\line(1,0){35}}
\put(50,45){\line(1,0){35}}
\put(100,45){\line(1,0){35}}
\put(150,45){\line(1,0){10}}

\put(10,10){\framebox(15,25){16}}
\put(35,25){\framebox(15,25){28}}
\put(60,10){\framebox(15,25){37}}
\put(85,25){\framebox(15,25){46}}
\put(110,10){\framebox(15,25){55}}
\put(135,25){\framebox(15,25){64}}

\end{picture}

We find an interesting regularity in how the U(8)
operation is built up by this sequence of gates, which is summarized by the
``dimensionalities" shown in the diagram.  The first U(4) operation has
$4^2=16$ free angle parameters; this is the dimensionality of the space
accessible with a single gate, as indicated.  With the second gate, this
dimensionality increases
only by 12, to 28.  It does not double to 32, for two reasons.
First, there is a single global phase shared by the two gates.  Second, there
is a set of operations acting only on the bit shared by the two gates, which
accounts for the additional reduction of 3.  Formally, this is summarized
by noting that 12 is the dimension of the {\it coset space} SU(4)/SU(2).
The action of the third gate increases the dimensionality by
another $9=16-1-3-3$.  9 is the dimension of the coset space
SU(4)/SU(2)$\times$SU(2).
The further subtraction by 3 results from
the duplication of one-bit operations on both bits of the added gate.  At
this point the dimensionality increases by nine for each succeeding
gate, until
the dimensionality reaches exactly 64, the dimension of U(8), at the sixth
gate.  In preliminary tests on four-bit operations, we found that the same
rules for the increase of dimensionality applied.  This permits us to make
a conjecture, just based on dimension counting, of
a lower bound on the number of two-bit
gates required to produce an arbitrary $n$-bit unitary transformation:
$\Omega(n)=\frac{1}{9}4^n-\frac{1}{3}n-\frac{1}{9}$.
It is clear that ``almost all" unitary
transformations will be computationally uninteresting, since they will
require
exponentially many operations to implement.

Finally, we mention that by combining the quantum gate constructions
introduced
here with the decomposition formulas for unitary matrices as used by Reck
{\it al.}\cite{Reck}, an explicit, exact simulation of {\it any} unitary
operator on $n$ bits can be constructed using a finite
number ($\Theta(n^3 4^n)$)
of two-bit gates, and using {\it no} work bits.  In outline, the procedure
is as follows: Reck {\it et al.}\cite{Reck} note that a formula exists
for the decomposition of any unitary matrix into matrices only involving
a U(2) operation acting in the space of pairs of {\it states} (not
{\it bits}):
$$U=(\prod_{x1,x2 \in \{0,1\}^m,\ x1>x2}T(x1,x2))\cdot D.$$
$T(x1,x2)$ performs a U(2) rotation involving the two basis states
$x1$ and $x2$, and leaves all other states unchanged; $D$ is a diagonal
matrix involving only phase factors, and thus can also be thought of as
a product of $2^{n-1}$ matrices which perform rotations in two-dimensional
subspaces.  Using the methods introduced above, each $T(x1,x2)$ can be
simulated in polynomial time, as follows: write a grey code connecting
$x1$ and $x2$; for example, if $n=8$, $x1=00111010$, and $x2=00100111$:

\smallskip
\begin{tabular}{lll}
1&00111010&{\it x1}\\
2&00111011&\ \\
3&00111111&\ \\
4&00110111&\ \\
5&00100111&{\it x2}
\end{tabular}

\smallskip
\ni Operations involving adjacent steps in this grey code require a
simple modification of the $\wedge_{n-1}$ gates introduced earlier.  The
$(n-1)$ control bits which remain unchanged are not all 1 as in our earlier
constructions, but they can be made so temporarily by the appropriate use
of ``NOT" gates ($\wedge_0(\sigma_x)$) before and after the application
of the $\wedge_{n-1}$ operation.  Now, the desired $T(x1,x2)$ operation
is constructed as follows: first, permute states down through the grey
code, performing the permutations (1,2), (2,3), (3,4), ... ($m$-2,$m$-1).
These numbers refer to the grey code elements as in the table above,
where m, the number of elements in the grey code, is 5 in the example.
Each of these permutations is accomplished by a modified
$\wedge_{n-1}(\sigma_x)$.  Second, the desired U(2) rotation is performed
by applying a modified $\wedge_{n-1}(U)$ involving the states $(m-1)$ and
$(m)$.  Third, the permutations are undone in reverse order: ($m$-2,$m$-1),
($m$-3,$m$-2), ... (2,3), (1,2).

The number of basic operations to perform all these steps may be easily
estimated.  Each $T(x1,x2)$ involves $2m-3$ (modified) $\wedge_{n-1}$ gates,
each of which can be done in $\Theta(n^2)$ operations.  Since $m$, the
number of elements in the grey code sequence, cannot exceed $n+1$, the
number of operations to simulate $T(x1,x2)$ is $\Theta(n^3)$.  There
are $O(4^n)$ $T$'s in the product above, so the total number of basic
operations to simulate any $U(2^n)$ matrix exactly is $\Theta(n^34^n)$.
(The number of steps to simulate the $D$ matrix is smaller and does not
affect the count.)  So, we see that this strict upper bound differs only
by a polynomial factor (which likely can be made better than $n^3$) from
the expected lower bound quoted earlier, so this Reck procedure is
relatively ``efficient" (if something which scales exponentially
may be termed so).  A serious problem with this procedure is that it
is extremely unlikely, so far as we can tell, to provide a polynomial-time
simulation of those special $U(2^n)$ which permit it, which of course
are exactly the ones which are of most interest in quantum computation.
It still remains to find a truly efficient and useful design
methodology for quantum gate construction.

\section*{Acknowledgments}
We are very grateful to H.-F. Chau, D. Coppersmith, D. Deutsch, A. Ekert
and T. Toffoli for
helpful discussions.
We are also most happy to thank the Institute
for Scientific Interchange in Torino, and its director Professor Mario
Rasetti, for making possible the workshop on quantum computation
in October, 1994,
at which much of this work was performed. A. B. thanks the financial
support of the Berrow's fund in Lincoln College (Oxford).

\pagebreak[1]

\samepage


\begin{thebibliography}{99}
\bibitem{foot1} The only non-trivial one-bit classical invertible operation is
${\scriptsize
\left(
\begin{array}{cc}
0 & 1 \\
1 & 0
\end{array}
\right)
}$.
\hihi
\end{thebibliography}
\end{document}